\documentclass[useAMS,fleqn,usenatbib]{mnras}
\usepackage[utf8]{inputenc}
\usepackage{amssymb,amsmath}
\usepackage{graphicx}
\usepackage[dvipsnames]{xcolor}

\usepackage{newtxtext,newtxmath}

\usepackage{array}
\newcolumntype{P}[1]{>{\centering\arraybackslash}p{#1}}

\title[ Non-parametric Reconstruction of $f_{\rm esc}$]{Non-parametric Reconstruction of Photon Escape Fraction from Reionization}

\author[Mitra \& Chatterjee]{
Sourav Mitra$^{1}$\thanks{E-mail:hisourav@gmail.com},
Atrideb Chatterjee$^{2}$
\\
$^{1}$ Department of Physics, Surendranath College, 24/2 M. G. Road, Kolkata 700009, India\\
$^{2}$Inter-University Centre for Astronomy and Astrophysics, Post Bag 4, Ganeshkhind, Pune 411007, India\\
}

\date{Accepted XXX. Received YYY; in original form ZZZ}

\pubyear{2023}

\begin{document}
\def\cyan{\color{cyan}}
\def\red{\color{red}}
\def\blu{\color{blue}}
\def\refjnl#1{{\rm#1}}
\def\aj{\refjnl{AJ}}                   
\def\actaa{\refjnl{Acta Astron.}}      
\def\araa{\refjnl{ARA\&A}}             
\def\apj{\refjnl{ApJ}}                 
\def\apjl{\refjnl{ApJ}}                
\def\apjs{\refjnl{ApJS}}               
\def\ao{\refjnl{Appl.~Opt.}}           
\def\apss{\refjnl{Ap\&SS}}             
\def\aap{\refjnl{A\&A}}                
\def\aapr{\refjnl{A\&A~Rev.}}          
\def\aaps{\refjnl{A\&AS}}              
\def\azh{\refjnl{AZh}}                 
\def\baas{\refjnl{BAAS}}               
\def\bac{\refjnl{Bull. astr. Inst. Czechosl.}} 
\def\caa{\refjnl{Chinese Astron. Astrophys.}} 
\def\cjaa{\refjnl{Chinese J. Astron. Astrophys.}} 
\def\icarus{\refjnl{Icarus}}           
\def\jcap{\refjnl{J. Cosmology Astropart. Phys.}} 
\def\jrasc{\refjnl{JRASC}}             
\def\memras{\refjnl{MmRAS}}            
\def\mnras{\refjnl{MNRAS}}             
\def\na{\refjnl{New A}}                
\def\nar{\refjnl{New A Rev.}}          
\def\pra{\refjnl{Phys.~Rev.~A}}        
\def\prb{\refjnl{Phys.~Rev.~B}}        
\def\prd{\refjnl{Phys.~Rev.~D}}        
\def\pre{\refjnl{Phys.~Rev.~E}}        
\def\prl{\refjnl{Phys.~Rev.~Lett.}}    
\def\pasa{\refjnl{PASA}}               
\def\pasp{\refjnl{PASP}}               
\def\pasj{\refjnl{PASJ}}               
\def\rmxaa{\refjnl{Rev. Mexicana Astron. Astrofis.}} 
\def\qjras{\refjnl{QJRAS}}             
\def\skytel{\refjnl{S\&T}}             
\def\solphys{\refjnl{Sol.~Phys.}}      
\def\sovast{\refjnl{Soviet~Ast.}}      
\def\ssr{\refjnl{Space~Sci.~Rev.}}     
\def\zap{\refjnl{ZAp}}                 
\def\nat{\refjnl{Nature}}              
\def\iaucirc{\refjnl{IAU~Circ.}}       
\def\aplett{\refjnl{Astrophys.~Lett.}} 
\def\apspr{\refjnl{Astrophys.~Space~Phys.~Res.}} 
\def\bain{\refjnl{Bull.~Astron.~Inst.~Netherlands}}  
\def\fcp{\refjnl{Fund.~Cosmic~Phys.}}  
\def\gca{\refjnl{Geochim.~Cosmochim.~Acta}}   
\def\grl{\refjnl{Geophys.~Res.~Lett.}} 
\def\jcp{\refjnl{J.~Chem.~Phys.}}      
\def\jgr{\refjnl{J.~Geophys.~Res.}}    
\def\jqsrt{\refjnl{J.~Quant.~Spec.~Radiat.~Transf.}} 
\def\memsai{\refjnl{Mem.~Soc.~Astron.~Italiana}} 
\def\nphysa{\refjnl{Nucl.~Phys.~A}}   
\def\physrep{\refjnl{Phys.~Rep.}}   
\def\physscr{\refjnl{Phys.~Scr}}   
\def\planss{\refjnl{Planet.~Space~Sci.}}   
\def\procspie{\refjnl{Proc.~SPIE}}   
\let\astap=\aap
\let\apjlett=\apjl
\let\apjsupp=\apjs
\let\applopt=\ao

\def\der{{\rm d}}
\label{firstpage}
\pagerange{\pageref{firstpage}--\pageref{lastpage}}
\maketitle

\begin{abstract}
One of the most crucial yet poorly constrained parameters in modelling the ionizing emissivity is the escape fraction of photons from star-forming galaxies. Several theoretical and observational studies have been conducted over the past few years, but consensus regarding its redshift evolution has yet to be achieved. We present here the first non-parametric reconstruction of this parameter as a function of redshift from a data-driven reionization model using a Gaussian Process Regression method. Our finding suggests a mild redshift evolution of escape fraction with a mean value of $4\%,7\%,\sim10\%$ at $z=2,6,12$. However, a constant escape fraction of $6-10\%$ at $z\gtrsim 6$ is still allowed by current data and also matches other reionization-related observations. With the detection of fainter high redshift galaxies from upcoming observations of JWST, the approach presented here will be a robust tool to put the most stringent constraint on escape fraction as well as reionization histories.
\end{abstract}


\section{Introduction} \label{sec:intro}

Cosmic reionization is one of the crucial epochs in the evolution of our Universe when neutral hydrogen
in the Inter-Galactic Medium (IGM) got ionized by the ultraviolet (UV) radiation from the  
first stars \citep{2009CSci...97..841C, 2018PhR...780....1D}. Currently available observational data of high-redshift quasars and measurements of cosmic microwave background (CMB) anisotropies suggest this epoch to take place somewhere between redshift 
$z \approx 6-20$ \citep{2011ApJS..192...18K,2016A&A...594A..13P,2020A&A...641A...5P, 2001AJ....122.2850B,2006AJ....132..117F,2015MNRAS.447..499M}.
Over the past few years, several analytical and numerical models have been proposed
to interpret these observations and put tighter limits on the epoch of reionization (EoR) \citep{2005MNRAS.361..577C,2011MNRAS.413.1569M,2011PhRvD..84l3522P,2012MNRAS.419.1480M,2014ApJ...793...29G,2015MNRAS.454L..76M,2017JCAP...11..028H,2019MNRAS.485L..24K,2021MNRAS.507.2405C,2022MNRAS.515..617M}.

The star-forming galaxies at higher redshifts are often thought to be dominant sources of ionizing photons (at least for $z\gtrsim6$). Not all the photons from those galaxies manage to {\it escape} into the IGM; only a fraction of it can, and this is known as escape fraction $f_{\rm esc}$. It is perhaps the most crucial parameter in understanding reionization and the early Universe.
Although a substantial number of observational attempts have been made to measure this quantity directly (see  \citealt{2022ARA&A..60..121R} for review), most of them are only limited to low-redshifts ($z \lesssim 4$). This is mainly because the IGM optical depth becomes very large at higher redshifts, especially in the time of reionization, making it impossible for direct measurement of $f_{\rm esc}$ \citep{2014MNRAS.442.1805I}.

Unfortunately, it is yet poorly constrained from the theoretical fronts as well \citep{2011ApJ...731...20F,2015MNRAS.453..960M,2019MNRAS.489.2669M}, and thus often parametrized as constant \citep{2015ApJ...811..140B,2015ApJ...802L..19R}
or having an ad-hoc functional form \citep{2012ApJ...746..125H} in the reionization models\footnote{However, more sophisticated ways of determining the redshift evolution of $f_{\rm esc}$ have been put forward recently by calibrating it with different astrophysical observations, see e.g., \citep{2019MNRAS.485L..24K, 2019ApJ...879...36F, 2021ApJ...917L..37C, 2023MNRAS.518..270K}.}.
However, the validity of those simple models is somewhat questionable. Although several independent
efforts have been put forward to predict its functional dependencies with redshifts,
they all lead to inconclusive and conflicting trends with a value ranging from 0.01 to 1.
In fact, in literature, one can find escape fractions are decreasing \citep{2000ApJ...545...86W,2014ApJ...788..121K,2020ApJ...892..109N} or increasing \citep{2010ApJ...710.1239R,2013MNRAS.431.2826F,2018MNRAS.480.2628F,2020MNRAS.496.4342L,2020MNRAS.496.4574Y,2022arXiv221206177T} or even non-evolving \citep{2008ApJ...673L...1G,2011MNRAS.412..411Y,2015MNRAS.453..960M,2016MNRAS.457.4051K} towards higher redshifts. All these contradictory results motivate us to re-investigate
the behaviour of this parameter as a function of redshift from a model-independent approach
in the context of current constraints on reionization.

In this {\it letter}, we present a robust technique for reconstruction of redshift evolution of escape fraction using Gaussian Process Regression (GPR) method. Gaussian Process (GP) has often been used in literature for the reconstruction of several estimators related to cosmology -- e.g., dark energy equation of state \citep{2010PhRvD..82j3502H,2011PhRvD..84h3501H,2012JCAP...06..036S,2019JCAP...07..042G}, Hubble parameter and deceleration parameter \citep{2012PhRvD..85l3530S}, the test of cosmological models with CMB data \citep{2017JCAP...09..031A}, separate 21-cm global signal from foreground contamination \citep{2018MNRAS.478.3640M} and  UV luminosity density parameter in the
context of reionization \citep{2021ApJ...922...95K}. Here, for the first time, we employ GP to reconstruct $f_{\rm esc}$ as a function of redshift by matching the available reionization
observables. This non-parametric methodology not only matches the current reionization
constraints but also offers a novel tool to check the efficacy of the simple parametric
models for escape fraction. We explain our approach in detail in the next section.

\section{Formalism} \label{sec:formalism}

To study the reionization of IGM, we use a data-constrained semi-analytical
reionization model described in \cite{2021MNRAS.507.2405C} which is based on \cite{2005MNRAS.361..577C,2006MNRAS.371L..55C}.
The model computes the ionization histories of both hydrogen and helium 
by solving a set of ordinary differential equations governing the evolution of
volume filling factor of individual ionized regions ($Q_i$).
Since we know from observations that there are high-density regions in the IGM that remain neutral even after reionization is completed, we need to consider IGM as an inhomogeneous medium. This is done by following an analytical prescription given by  \cite{2000ApJ...530....1M} where reionization is said to be complete (i.e. $Q_i$ becomes 1) once all the low-density regions with overdensities $\Delta_i<\Delta_{\rm c}$ are ionized, where $\Delta_{\rm c}$ is the critical density which is computed from the mean separation of the ionizing sources. In this work, it has been taken as constant at $\sim 60$. However, varying this parameter won't affect the overall reionization history much as shown in \cite{2018MNRAS.473.1416M}. In this formalism, the mean free path of ionizing photons is determined from high-density regions of the IGM as
\begin{equation}
    \lambda_{\rm mfp}(z) = \frac{\lambda_0}{[1-F_V(z)]^{2/3}}
\end{equation}
where $F_V$ is the fraction of volume which is ionized or, equivalently, the fraction
filled up by the low-density regions. Given the density probability distribution ({\it lognormal} at low densities, changing to a {\it power-law} at high densities), it is straightforward to calculate this quantity \citep{2005MNRAS.361..577C}. $\lambda_0$ is a normalization factor and treated as a free parameter in this model, which can be fixed by comparing with low redshift observations, as we shall see later.

The major uncertainty in any reionization modelling is identifying the sources of ionizing
photons. In this work, we assume those to be quasars (dominant sources for $z<6$ and thus key contributors for helium reionization) and high-redshift stars. The quasar contribution
can be calculated by computing their ionizing emissivities from the observed
quasar luminosity function (LF) at $z<7.5$ \citep{2007ApJ...654..731H,2019MNRAS.488.1035K} and modelling those accurately is not so critical as they probably do not dominate the photon budget at higher redshifts. On the other hand, calculating the stellar contribution is non-trivial as it involves various complicated and unknown physical processes. For our model, we neglected the contribution of metal-free Population III stars and only considered the contribution of PopII stars, which is justified as shown in \cite{2021MNRAS.507.2405C}. The model also incorporates the suppression of star formation in low-mass haloes (known as {\it radiative feedback}) self-consistently from the thermal evolution of IGM. 

The ionizing photon production rate from stellar sources is then calculated as
\begin{equation}
    \dot{n}_{\rm ph, \rm stellar}(z)= \rho_b f_{*} f_{\mathrm{esc}}(z) \frac{\der f_{\mathrm{coll}}}{\der t} \int^{\infty}_{\nu_H}  \left(\frac{\der N_{\nu}}{\der M} \right) \der \nu
\end{equation}
where $\rho_b$ is the mean comoving density of baryons in the IGM, $\der f_{\mathrm{coll}}/\der t$ is the rate of collapse fraction of the dark matter halos, $\nu_{H}$ is the threshold frequency for hydrogen photoionization, and $f_*$ is the star formation efficiency. The quantity $\der N_{\nu}/ \der M$, the number of photons emitted per frequency range per unit mass of the star, depends on the stellar spectra and initial mass function (IMF) of the stars \citep{2005MNRAS.361..577C}. Using a standard Salpeter IMF in the mass range $1-100 M_{\odot}$ with a metallicity of $0.05 M_{\odot}$, $\der N_{\nu}/ \der M$ has been computed from the stellar synthesis models of \cite{2003MNRAS.344.1000B}. Although both $f_*$ and $f_{\rm esc}$ can, in principle, be a function of redshift and halo masses, we neglect their mass dependencies to keep our analysis simple. Furthermore, we find $f_*$ to be roughly constant at $\sim 0.04$ with redshift by matching the observed UV luminosity function at $6\le z \le 10$ \citep{2021AJ....162...47B}, which is consistent with the earlier works of \cite{2015MNRAS.454L..76M,2018MNRAS.479.4566M}.

Finally, the key quantity of our interest, the escape fraction as a function of redshift, deserves a thorough investigation. For simplicity, this parameter is often assumed to be either constant \citep{2015ApJ...811..140B,2015ApJ...802L..19R}
or have an impromptu parametric dependency with redshifts \citep{2012ApJ...746..125H} -- both of these are somewhat unconvincing. Also, in most cases, the parametrizations are not flexible enough to track the underlying true behaviour of that quantity. Thus, in this work, rather than taking any particular functional form, we take the globally-averaged $f_{\rm esc}(z)$ as a non-parametric function of redshift and reconstruct it by matching the reionization model with current observations using GPR.

\subsection{\label{sec:GPR}Non-parametric reconstruction}
We start with discretizing the quantity $f_{\rm esc}(z)$ 
into several equally spaced
redshift nodes $z_i$ ranging $2\leq z\leq 20$ \footnote{From the current observations, we do not have any significant data points (related to reionization) beyond this redshift range, and therefore the reconstruction would be very poor at values beyond this range.}. The value of $f_{\rm esc} (z_i)$ in each node is considered as a free
parameter (say $\epsilon_i$) in this analysis. We then use those parameters as a set of training inputs and employ a  GPR algorithm. GP is particularly useful for reconstructing a function from data without requiring its parametrization. For a reconstructed function $f_{\rm esc}(z)$, GP is often characterized by a mean $\mu(z)$ and a covariance {\it kernel} function ${\rm cov}(f_{\rm esc}(z), f_{\rm esc}(z')) = k(z, z')$ which is defined by a small number of hyperparameters \citep{books/lib/RasmussenW06}. These hyperparameters are optimized consistently according to the data in each GP realisation. Although there are ample options for possible kernel functions, we have chosen the Radial Basis Function (RBF) defined as $k(z, z') = \sigma^2 \exp\left(-\frac{(z-z')^2}{2l^2}\right)$ with two hyperparameters -- signal variance $\sigma^2$ and lengthscale $l$. The main advantage of using this kernel is that it is infinitely differentiable, ensuring a smooth reconstruction of $f_{\rm esc}(z)$. We do not explore the possibility of other suitable kernels here but leave that as an extension of the current work. In general, the mean function $\mu(z)$ should be chosen such that it is close to the {\it true} $f_{\rm esc}(z)$. However, the mean will be adjusted through the spread $\sigma^2$ for each GP realization during the analysis, even though we have fixed its initial value. This gives us the flexibility to choose $\mu(z)$. We have tested that our results remain the same for any choice of its fiducial value as long as it is physically consistent, that is, between $0$ and $1$ (from the definition of $f_{\rm esc}$).

\subsection{\label{sec:data}Dataset and likelihood}
Although the reionization model can be constrained by comparing with a wide variety of observational data, in this work, we use only those most effective to the reconstruction of $f_{\rm esc} (z)$ and $\lambda_0$. To compute the likelihood function, we use: (i) measurements of photoionization rate $\Gamma_{\rm PI}$\footnote{Calculating $\Gamma_{\rm PI}(z)$ requires frequency-dependent HI absorption cross-section $\sigma_{\rm HI}(\nu)$ and mean free path for hydrogen ionizing photons $\lambda_{\rm HI}(z;\nu)$. We refer the reader to follow \cite{2005MNRAS.361..577C} for detailed calculations and the assumptions involved therein.}
at $2.4\le z\le 6$ obtained using joint analysis of quasar absorption spectra and a large set of hydrodynamical simulations \citep{2011MNRAS.412.2543C,2013MNRAS.436.1023B,2018MNRAS.473..560D}, (ii) redshift evolution
of Lyman Limit Systems (LLS), ${\rm d}N_{\rm LL}/{\rm d}z$ in the range $2.3\le z\le 5.8$ \citep{2010ApJ...721.1448S,2010ApJ...718..392P,2019MNRAS.482.1456C} (inclusion of this dataset
is essential in constraining the parameter $\lambda_0$) and (iii) electron scattering optical depth $\tau_{\rm el}$ from recent Planck 2018 data \citep{2020A&A...641A...5P}. We further set a prior on the neutral hydrogen fraction $x_{\rm HI}$ (defined as the ratio of a number of neutral hydrogen to total hydrogen) using model-independent upper limit obtained from the effective Ly$\alpha$ optical depth data \citep{2015MNRAS.447..499M}.

In principle, the choice for the total number of redshift nodes can depend on the sensitivity of data on the parameter $f_{\rm esc}(z)$. However, we thoroughly investigate that we can achieve a fair reconstruction by considering a minimum of 4 redshift nodes. Including a higher number of nodes rather instigates more computational time than adequately improving reconstruction quality. Thus, the model has 5 free parameters: four GP parameters  $\epsilon_{1}$, $\epsilon_{2}$, $\epsilon_{3}$, $\epsilon_{4}$ corresponding to the redshifts 3, 6, 9, 12 respectively, for the reconstruction of the escape fraction $f_{\rm esc}(z)$, and one reionization model parameter $\lambda_{0}$. 

Given the set of GP free parameters $\{\epsilon_1, \epsilon_2, \epsilon_3, \epsilon_4\}$, we use a publicly available \texttt{Python} package called
\texttt{scikit-learn} to perform the GPR and generate the reconstructed $f_{\rm esc}(z)$. Finally, comparing the predictions of our model with the observations mentioned before, we constrain all the free parameters.
For that, we employ a  Markov Chain Monte Carlo (MCMC) based likelihood analysis using our previously developed parameter estimation package \texttt{CosmoReionMC} \citep{2021MNRAS.507.2405C}.
We define our likelihood function as given below,
\begin{equation}
\mathcal{L}_{Re}=\frac{1}{2}\sum^{N_{\mathrm{obs}}}_{\alpha=1}\left[\frac{\zeta^{\mathrm{obs}}_{\alpha}-\zeta^{\mathrm{th}}_{\alpha}}{\sigma_{\alpha}}\right]^2.
\label{eqn:Likelihood}
\end{equation}
While $\zeta^{\mathrm{obs}}_{\alpha}$ represents the set of $N_{\mathrm{obs}}$ observational data associated with photoionization rates and the redshift distribution of the Lyman-Limit system, $\zeta^{\mathrm{th}}_{\alpha}$ represents the values from the theoretical model. The $\sigma_{\alpha}$ are the observational error bars.

We assume broad flat priors for all the five free parameters. The MCMC chains are run with a sufficient number of walkers and steps until the desired convergence criterion is satisfied. Throughout the analysis, we assume a flat $\Lambda$CDM model with the cosmological parameters fixed at their best-fit values as given in Planck 2018 data \citep{2020A&A...641A...5P}.

\section{Results} \label{sec:results}

\begin{figure}
\centering
\includegraphics[width=0.95\columnwidth]{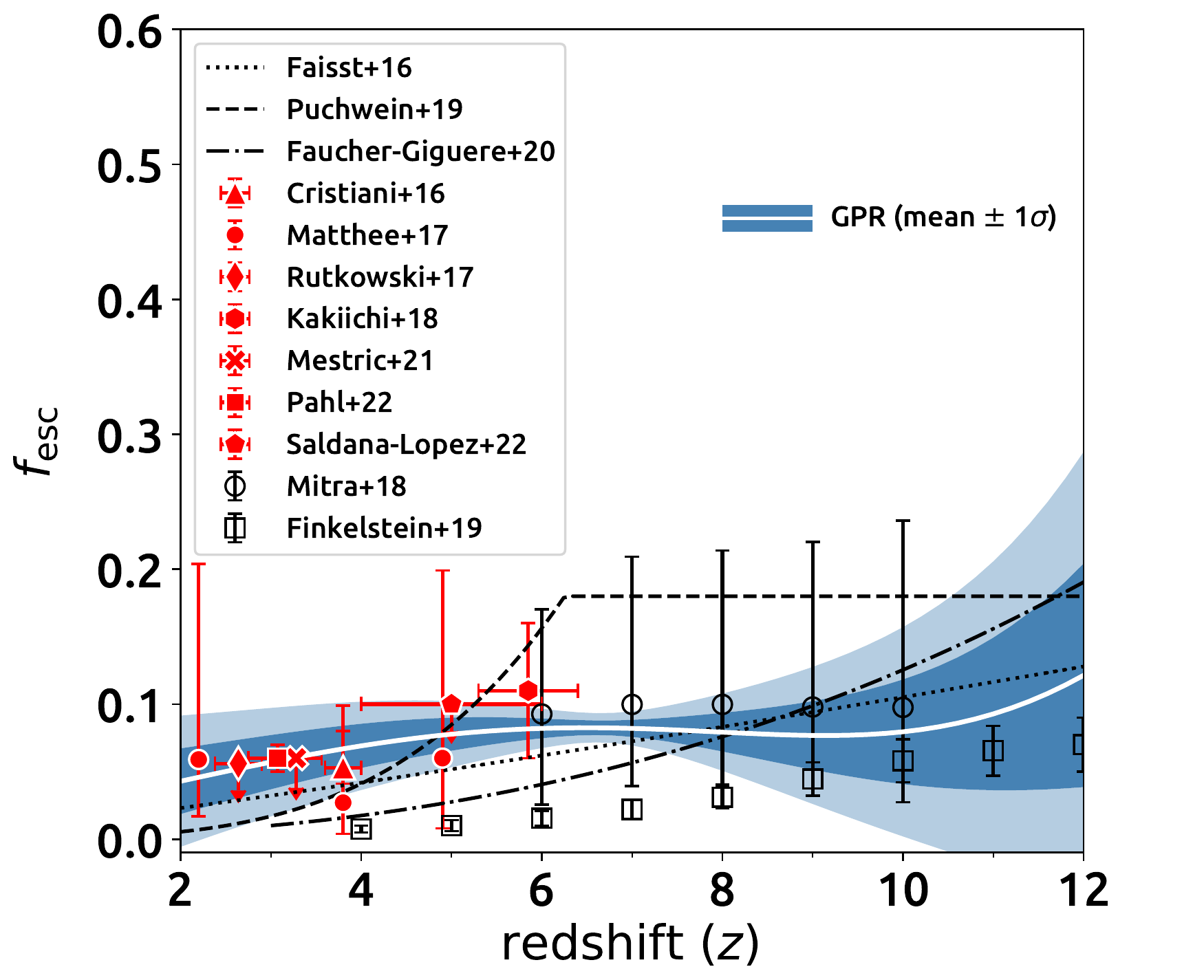}
\caption{Non-parametric reconstruction of mean $f_{\rm esc}(z)$ (solid white line) with 1-$\sigma$ (dark blue shaded) and 2-$\sigma$ (light blue shaded) confidence limits. Red data points are from different observations and surveys \citep{2016MNRAS.462.2478C,2017MNRAS.465.3637M,2017ApJ...841L..27R, 2018MNRAS.479...43K,
2021MNRAS.508.4443M,
2021MNRAS.505.2447P, 
2022A&A...663A..59S} whereas the black points are taken from the recent findings of \citet{2019ApJ...879...36F} (open squares) and \citet{2018MNRAS.479.4566M} (open circles). For comparison, we also plot some popular fitting functions by different line styles \citep{2016ApJ...829...99F, 2019MNRAS.485...47P, 2000ApJ...530....1M, 2020MNRAS.493.1614F}. }
\label{fig:fesc}
\end{figure}

In Figure \ref{fig:fesc}, we show the redshift evolution of our reconstructed globally-averaged $f_{\rm esc}(z)$. The solid white line corresponds to the mean model, while the dark and light blue shaded regions signify the 1 and 2-$\sigma$ confidence limits, respectively. The error bars seem to be larger at higher redshifts due to the lack of current reionization observables, and that is something which can be improved significantly from future probes \citep{2022ARA&A..60..121R}. We find a mildly increasing mean escape fraction with an increase of redshift, $f_{\rm esc}=4\%,7\%,\sim10\%$ at $z=2,6,12$. However, considering the 2-$\sigma$ limits, there could be a maximum increase of $f_{\rm esc}\sim30\%$ at $z=12$. Still, a constant escape fraction of 6-10\% at $z\gtrsim 6$\footnote{This hints that a mechanism other than evolving $f_{\rm esc}$ would be needed to explain the non-monotonous evolution of cosmic emissivity seen in \cite{2021MNRAS.507.6108O} during this epoch.} is allowed by the current data, which essentially suggests a reionization scenario driven by single stellar population (PopII). This agrees with the previous works by  \cite{2015MNRAS.454L..76M,2018MNRAS.479.4566M} where a Principal Component Analysis (PCA) has been implemented based on the same reionization model we use here. They obtained a non-evolving escape fraction of $\sim10\%$ in the redshift ranges $z=6-10$ (black open circles in Figure \ref{fig:fesc}). Although the PCA-based model also follows a non-parametric reconstruction approach, it somewhat depends on the chosen fiducial model and requires a significantly higher number of free parameters (PCA modes) in the MCMC steps. A similar trend has also been reported by \cite{2016MNRAS.457.4051K} where they found a steep rise in $f_{\rm esc}$ between redshifts $3.5$ to $5.5$ and then becoming constant at 14-20\% during reionization epoch.

We also compare our results with other recent determinations. The red solid points with errorbars are taken from different observations and surveys \citep{2016MNRAS.462.2478C,2017MNRAS.465.3637M,2017ApJ...841L..27R,2018MNRAS.479...43K,2021MNRAS.508.4443M,2021MNRAS.505.2447P,2022A&A...663A..59S}. Our reconstructed value suggests a smaller $f_{\rm esc}\lesssim 10\%$ (at 2-$\sigma$) at lower redshifts which is in excellent agreement with almost all of these observations. Note that we did not use any of these data while constraining the $f_{\rm esc}$. Perhaps, a more promising match can be seen with the prediction of \cite{2016ApJ...829...99F} (black dotted lines in the figure), where they provided the first observational inference of $f_{\rm esc}$ in galaxies at $z > 6$ based on the H$\alpha$ observations. For comparison, we show some of the well-known parametric forms of escape fraction as a function of redshift by the dashed lines from \cite{2019MNRAS.485...47P} and dot-dashed lines from \cite{2020MNRAS.493.1614F}. The black open points with errorbars are taken from the recent finding of \cite{2019ApJ...879...36F} where they obtained a significantly low ($<5\%$) and mildly increasing escape fraction with an increase of redshift in order to reionize the Universe. Their results are very similar to those recently achieved from \texttt{SPHINX} suite of simulations \citep{2022MNRAS.515.2386R}. Although our GP reconstructed escape fraction is in broad agreement with most of those, the disagreement is more pronounced with the models where a very high escape fraction at low redshifts or a rapidly increasing $f_{\rm esc} (z)$ is needed \citep{2012ApJ...746..125H,2018MNRAS.480.2628F,2020ApJ...892..109N}. We should mention here that, while we have reconstructed $f_{\rm esc}$ up to redshift $20$, we did not show the values beyond $z=12$ as the errorbars associated with the GP become significantly large (ranging from 0 to 100\% at 2-$\sigma$ level) due to the non-availability of data during this epoch. 

\begin{figure}
\centering
\includegraphics[width=0.97\columnwidth]{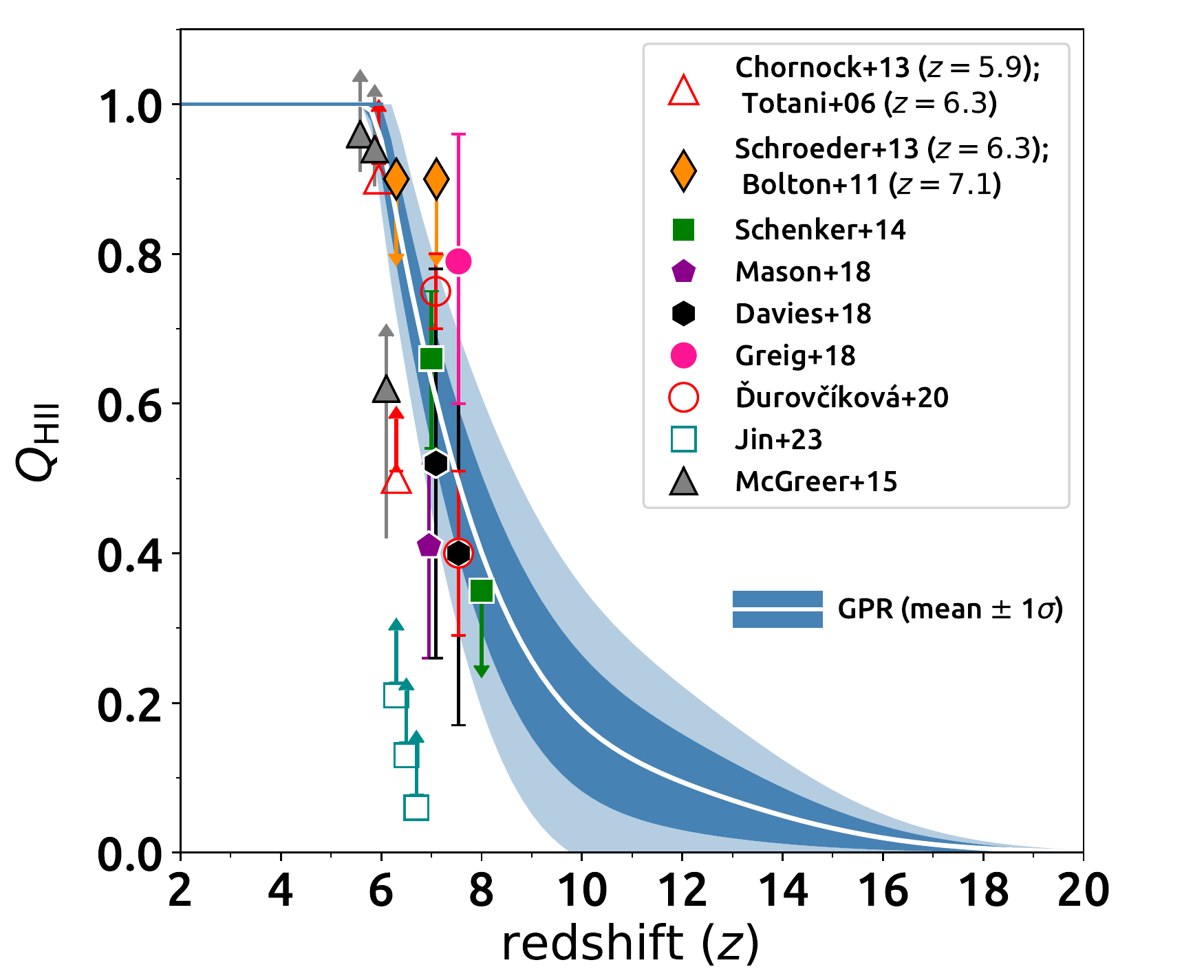}
\caption{Redshift evolution of volume-filling factor of ionized hydrogen $Q_{\mathrm{HII}}(z)$ with 1 and 2-$\sigma$ confidence limits (colour-coded) obtained from the GP+MCMC analysis. The points with error bars taken from various existing observations (as listed here and further in the text) are shown for comparison.}
\label{fig:QHII}
\end{figure}

Figure~\ref{fig:QHII} shows our reconstructed reionization history, namely the redshift evolution of the volume-filling factor of ionized ($\mathrm{HII}$) regions $Q_{\mathrm{HII}}(z)$. Reionization is said to be completed when $Q_{\mathrm{HII}}$ becomes 1, and that happens around $5.8\lesssim z\lesssim 6.3$ (2-$\sigma$ limits). We also find that the Universe was $50\%$ ionized ($Q_{\mathrm{HII}}=0.5$) at $6.9\lesssim z\lesssim 9.2$ (2-$\sigma$ limits). For comparison, we have shown various observational constraints available in the literature, such as the measurements from GRB (Gamma-Ray Burst) host galaxies (open red triangles from \citealt{2006PASJ...58..485T,2013ApJ...774...26C}), quasar near zones observations at $z\sim 6-7$ (orange diamonds from \citealt{2011MNRAS.416L..70B,2013MNRAS.428.3058S}), measurements of Lyman-$\alpha$ emission at redshifts $\sim7-8$ (filled green squares from \citealt{2014ApJ...795...20S} and purple pentagon from \citealt{2018ApJ...856....2M}), damping wing analysis of highest redshift ($z>7$) quasars (black hexagons from \citealt{2018ApJ...864..142D}, filled pink circle from \citealt{2019MNRAS.484.5094G} and open red circles from \citealt{2020MNRAS.493.4256D}) and dark pixel analysis of $z\sim6$ quasar spectra (open cyan squares from \citealt{2023ApJ...942...59J} and filled grey triangles from \citealt{2015MNRAS.447..499M}). Considering the fact that we did not include these datasets in our MCMC analysis (except a prior on $x_{\mathrm{HI}}$ forcing reionization to be completed by $z\approx6$), the match of our model prediction with those seems to be quite impressive. 

One can see that the current data still permit a wide range of reionization scenarios at $z\gtrsim 8$. A similar trend can also be found in the $f_{\rm esc} (z)$ plot. This is expected as the high-redshift constraints in this analysis solely depend on the value of $\tau_{\rm el}$ coming from CMB observations and thus remain somewhat weaker. One can, in principle, include some of these observed constraints on $Q_{\mathrm{HII}}$ at redshift $\sim7-8$ and get a significantly tighter reionization history (see, e.g. \citealt{2018MNRAS.479.4566M}), but keep it in mind that most of these observational results are highly model-dependent and also might get changed in future. For this reason, we refrain from using those in our likelihood function. Those regions will be extensively constrained with the upcoming probe like the James Webb Space Telescope (JWST), which aims to detect more and fainter galaxies during the epoch of reionization \citep{2022ARA&A..60..121R}.

\section{Conclusions}
 This letter is the first attempt to reconstruct the escape fraction ($f_{\rm esc}$) of ionizing photons as a function of redshift using a Gaussian Process (GP). We employ a physically motivated, data-driven semi-analytic model constrained against a number of recent datasets related to hydrogen reionization using an MCMC-based likelihood analysis. We found that the observations related to reionization favour a mildly increasing $f_{\rm esc}(z)$ with redshift, from a mean value of $4\%$ at $z=2$ to $7\%$ at $z=6$ and $\sim 10\%$ at $z=12$, which is in good agreement with most of the current determinations (Figure \ref{fig:fesc}). Such a model can simultaneously match most of the available observed constraints on IGM neutral fraction (Figure \ref{fig:QHII}), as well as produce a reionization optical depth $\tau_{\rm el}$ of $0.056^{+0.012}_{-0.016}$ (2-$\sigma$ limits) consistent with the Planck 2018 measurements \citep{2020A&A...641A...5P}. On the other hand, models with rapidly increasing or a very high ($>10\%$) escape fraction at lower redshifts ($z<8$) are clearly in tension with our results.

 We should mention here a few caveats of the present analysis. For simplicity, we have assumed a constant stellar IMF throughout this work, but, in principle, the inferred $f_{\rm esc}$ might depend on this choice. Thus it would be interesting to investigate the degeneracy between $f_{\rm esc}$ and the stellar IMF. Furthermore, we assume a mass-independent $f_*$ to match the observed LF of galaxies at $6\le z\le 10$. Although the same technique can also be applied for low-redshift galaxies ($z\sim 3-5$), we note that this simple $f_*$ struggles to match those observables, especially at the brighter end of LFs hinting a possible mass-dependencies \citep{2018MNRAS.479.4566M} in the star formation efficiency parameter. Both of these shortcomings are left for future investigation. 
 
 However, the most interesting result is that a non-evolving $f_{\rm esc}$ of 6-10\% (at 2-$\sigma$ limits) at redshifts $z\gtrsim 6$ is well-permitted by the data. This essentially tells us that most of the ionizing photons can be contributed by a single stellar population (PopII) in order to reionize the IGM. {\it Specifically, a simplistic model with a slightly increasing $f_{\rm esc}$ at lower redshifts ($2\lesssim z\lesssim 6$) and 6-10\% (redshift-independent) afterwards would be more promising from both observational and theoretical fronts}. Perhaps to get more insights about reionization, one could use this simple model of $f_{\rm esc}$ in ab-initio numerical simulations and tune other parameters to match the observables, saving a considerable amount of computational time and resources.  
 Moreover, with the upcoming observations from JWST, the approach presented here will be a very powerful tool in order to discard some of the still-allowed models with higher $f_{\rm esc}$ or early reionization and shape our understanding of high-redshift Universe. 
\section{ACKNOWLEDGMENTS}

SM wishes to thank Dhiraj Kumar Hazra, Debabrata Adak and Aditi Krishak for valuable discussions on Gaussian Process. AC wishes to acknowledge the computing facility provided by IUCAA.

\section{Data Availability}
The observational datasets used here are taken from the literature, and the code used for this work will be shared on request with the corresponding author.
\bibliographystyle{mnras}
\bibliography{reion}

\end{document}